\documentclass[12pt]{iopart}
\usepackage{iopams,amssymb,graphicx,amsbsy,hyperref,color,CJK}
\begin{document}
\begin{CJK*}{UTF8}{mj}

\title{Multistability and Variations in Basin of Attraction in Power-grid Systems}

\author{Heetae Kim$^{1,2}$, Sang Hoon Lee$^{3}$, J\"orn Davidsen$^{4}$, and Seung-Woo Son$^{2,4,5}$
}

\address{$^1$Department of Industrial Engineering, University of Talca, Curi\`o, 3341717, Chile\\
$^2$Asia Pacific Center for Theoretical Physics, Pohang, 37673, Korea\\
$^3$Department of Liberal Arts, Gyeongnam National University of Science and Technology, Jinju, 52725, Korea \\
$^4$Department of Physics and Astronomy, University of Calgary, Calgary, Alberta T2N 1N4, Canada \\
$^5$Department of Applied Physics, Hanyang University, Ansan, 15588, Korea}
\eads{\mailto{hkim@utalca.cl},\mailto{lshlj82@gntech.ac.kr},\mailto{davidsen@phas.ucalgary.ca},\mailto{sonswoo@hanyang.ac.kr}}
\end{CJK*}
\vspace{10pt}
\begin{indented}
\item[]July 2018
\end{indented}

\begin{abstract}
Power grids sustain modern society by supplying electricity and thus their stability is a crucial factor for our civilization. 
The dynamic stability of a power grid is usually quantified by the probability of its nodes' recovery to phase synchronization of the alternating current it carries, in response to external perturbation.
Intuitively, the stability of nodes in power grids is supposed to become more robust as the coupling strength between the nodes increases.
However, we find a counterintuitive range of coupling strength values where the synchronization stability suddenly droops as the coupling strength increases, on a number of simple graph structures. 
Since power grids are designed to fulfill both local and long-range power demands, such simple graph structures or graphlets for local power transmission are indeed relevant in reality. 
We show that the observed nonmonotonic behavior is a consequence of transitions in multistability, which are related to changes in stability of the \emph{unsynchronized} states. Therefore, our findings suggest that a comprehensive understanding 
of changes in multistability are necessary to prevent the unexpected catastrophic instability in the building blocks of power grids. 
\end{abstract}

\vspace{2pc}

%\maketitle

\section{Introduction}
\label{sec:introduction}

The stable supply of electricity matters to a wide range of areas in our society today.
Even a small failure at a single transmission line can cause a massive cascade resulting in a large-scale blackout, which has motivated a number of recent studies focusing on the stability of power grids~\cite{Chaos.17.2,Chaos.19.1,Motter:2013iw,Dorfler:2013ew,Yang:2017io}. 
Some studies estimated structural stability of power grids in response to the amount of transmission load when the electricity flows from power plants to substations~\cite{Yang:2017io,Arianos:2009ua,PhysRevE.69.025103,PhysRevE.69.045104,EPJB.46.1,Chen2010595,Rohden:2017bu,Schafer:2017cc,Po:2017fp}, while
others focused on the dynamical synchronization stability of power-grid nodes in response to the disturbance in their phase or frequency~\cite{Menck:2012vc,Witthaut2012,Schultz2014detours,Menck:2014fn,Rohden:2014fy,Nishikawa:2015gl,BasinStability_NJP,Kim:2016kd}. The synchronization stability refers to the ability to recover synchronization of power-grid nodes after the nodes get perturbed. 
Considering the power-grid nodes as oscillators connected by transmission lines, the second-order Kuramoto-type model~\cite{Dorfler:2013ew,Nishikawa:2015gl,Filatrella:2008co,Ji:2014ina}---the so called swing equation in electric engineering---can describe the phase of the alternating current flowing in power grids.
The main question is whether the power-grid system retains synchronization after being perturbed or falls into desynchronization.
The volume of the basin of attraction to synchronization in the phase space of perturbation is interpreted as the probability of synchronization recovery, so-called ``basin stability", which has been widely used to quantify the synchronization stability of power grids~\cite{Schultz2014detours,Menck:2014fn,BasinStability_NJP,Kim:2016kd,Menck2013basin,vanKan:2016ie,Schultz:ch,Hellmann:2016js,Mitra:2017jp,Nitzbon:2017fo,Kittel:2017tz}.

Synchronization dynamics of power-grid nodes depends on the coupling strength between the nodes, which corresponds to the transmission capacity between the nodes composed of power plants (producers) and substations (consumers) in power grids. 
Therefore, the basin stability as a measure of the synchronization stability should also be intrinsically related to the coupling strength, e.g., weak coupling strength cannot be enough to mediate between nodes, causing low level of synchronization stability, whereas strong coupling strength generally ensures high level of stability~\cite{Menck2013basin,Menck:2012vc}.  
When the coupling strength is larger than a certain threshold, the basin stability of a node can even reach the unity, i.e., the node always recovers its synchrony against any given external perturbation in the phase space~\cite{Menck:2012vc}.

A concomitant result with this expectation is the case of an infinite busbar model. 
The infinite busbar model represents a power grid that includes an oscillatory power-grid node and a connected system as an environment, where the basin stability of power-grid nodes increases monotonically as the coupling strength increases (see section~\ref{sec:pattern} for details)~\cite{Menck:2014fn}. 
However, \cite{Kim:2016kd} 
revealed in small systems that the transition of basin stability shows various patterns including sudden drooping of the basin stability for the increased value of coupling strength.  
Although the basin stability is still useful even when the transition features anomalous patterns, and it is reliable even in the presence of fractal basin boundaries according to \cite{Schultz:ch}, the potential inconsistency should be clarified by uncovering the underlying mechanism in order to guarantee secure electricity supply. In this paper, we investigate in-depth the transition pattern of basin stability as a function of coupling strength
in the phase space of perturbation to power-grid nodes.

In this study, we find  
nontrivial behaviors of nodes in small-size networks that potentially cause the synchronization failure. 
First, we discover a nonmonotonic transition pattern characterized by the basin stability drooping even when the coupling strength increases. This implies that the stability found for small values of coupling strength does not necessarily guarantee 
the same level of stability for the larger values of coupling strength. Second, we reveal nontrivial fractal-like fine structures in the phase space of perturbation. Therefore, a small change in the perturbation can make a great difference in the final synchrony of a power-grid system, which is 
similar to the behavior observed in the case of fractal basin boundaries and riddled or intermingled basins of attraction~\cite{Schultz:ch}.
Even though we observe these phenomena in a small system, we emphasize that such a small graph structure or graphlet~\cite{Sarajlic:2016wx} is at the core of local power transmission and essentially plays the role of the fundamental building blocks of large power-grid systems.

This paper is organized as follows:
In section~\ref{sec:Dynamics}, we introduce the equation of motion that describes the synchronization behavior between oscillators in power-grid systems. We present our detailed results and the interpretation in section~\ref{sec:results}.
Finally, in section~\ref{sec:summary_and_discussions}, we present the concluding remarks and future outlook.

\section{Power-grid Dynamics}
\label{sec:Dynamics}
\subsection{Synchronization in Power Grids}
\label{subsec:Synchronization}
A power grid refers to an interconnected network of electric power transmission lines, whose node elements are composed of power plants and substations. 
In general substations are considered as \emph{consumers}, since substations are the gateways where electricity flows out to the consumers through distribution networks. 
The other group is composed of power plants, which plays the role of \emph{producers} of electricity in power grids.
We specify the power producers with positive values of net power input $P_i>0$ and the consumers as negative values $P_i<0$ ($|P_i|$ amount of power output) in the grid, for each node $i$, which will be used in the forthcoming equation describing the phase dynamics.

A power-grid node is an oscillator in terms of the phase angle of its alternating current that oscillates sinusoidally with a designated angular frequency.
The oscillation dynamics is described by the motion of a damped driven pendulum, the swing equation~\cite{Araposthatis:1981vt}, which is a Kuramoto-type model and it is widely used in power-grid studies~\cite{Motter:2013iw,Schultz2014detours,Menck:2014fn,Nishikawa:2015gl,Ji:2014ina,Menck2013basin,RevModPhys.77.137}. However, in contrast to the conventional (first-order) Kuramoto model where the system is completely governed by the first derivative of the phase angle~\cite{RevModPhys.77.137}, the swing equation involves the second derivative of the phase angle, or the time derivative of the angular frequency as presented in the following.

Applying the swing equation to power-grid nodes, the equation of motion for the phase angle $\theta_i$ of each node $i$ is given by~\cite{Filatrella:2008co}:
\begin{equation}
\ddot \theta_i = \dot \omega_i = P_i - \alpha_i \dot \theta_i - \sum_{i \ne j} K_{ij} a_{ij} \sin(\theta_i - \theta_j),
\label{eq:Kuramoto_type_equation}
\end{equation}
where the source term $P_i$ is the net-power input of node $i$ ($P_i > 0$ if the node $i$ is a producer and $P_i < 0$ for a consumer, as we introduced before---collectively, we denote the two cases by $P_p > 0$ and $P_c < 0$, respectively.); the damping term is the product of the dissipation coefficient $\alpha_i$ and the deviation of the angular frequency to the reference frame $\dot \theta_i$; $K_{ij}$ is the coupling strength that corresponds to the transmission capacity of the line between nodes $i$ and $j$; the adjacency matrix element, which describes the interaction or network structure, is given by $a_{ij} = 1$ if nodes $i$ and $j$ are connected, and $a_{ij} = 0$ otherwise. On top of the interaction substrate given by $a_{ij}$, the coupling between the phase angles of nodes $i$ and $j$ is given by the sine function, as in the conventional Kuramoto model~\cite{RevModPhys.77.137}. Note that the phase angle of node $i$ denoted by $\theta_i$ represents the phase difference, and $\omega_i = \dot \theta_i$ is the difference of the angular frequency of node $i$, from the desired constant angular frequency called \emph{rated frequency}; these reference points are set in order to ensure the vanishing mean value $\sum_i \dot \theta_i = \sum_i \omega_i = 0$. 

When power-grid nodes are synchronized in a steady state, the angular frequency is equal to the rated frequency, such that $\omega_i = 0$ for all $i$'s, which means that all of the nodes keep the rated frequency of the power system (\emph{frequency synchronization}). 
To characterize the stability of power-grid systems, we thoroughly investigate the response of nodes against the coupling strength change in terms of synchronization dynamics. In particular, we examine the detailed transition pattern of the basin stability observed in the basic units of power-grid systems composed of simple graph structures, which sometimes entail nonmonotonic pattern of sudden drooping.

For the sake of simplicity, we use the global coupling strength $K_{ij}=K$ as the control parameter adjusted from zero to the values large enough to guarantee the power grid's recovery to its original synchrony against the given range of perturbations.
We also set the dissipation coefficients $\alpha_i = \alpha = 0.1$ for all of the nodes~\cite{Schultz2014detours,Menck:2014fn,Menck2013basin}. For numerical integration of Eq.~(\ref{eq:Kuramoto_type_equation}), we use the fourth-order Runge-Kutta method~\cite{nr:07} and the convergence criteria~\cite{Schultz2014detours} of the deviation of the angular frequency from the rated frequency, $|\omega_i | < 5\times10^{-2}$ for every node $i$, which we believe is reasonable based on the observed fluctuations data.

\subsection{Synchronization Stability of Power Grids}
\label{sec:Power_Grid_Stability}
We use the numerical Monte Carlo approach~\cite{Menck2013basin} to estimate the basin stability $B$, our measure of node's ability to recover its synchrony.
More precisely, $B$ is the fraction of volume in the phase space composed of the phase and angular frequency disturbance as the external perturbation, from which the synchronization recovery occurs. 
In order to measure $B$, after the system reaches the synchrony, i.e., $\dot \theta_i = \omega_i = 0$ for all $i$, we perturb each node $j$ by imposing a new phase and angular frequency values $(\theta_j^*, \omega_j^*)$ chosen from the phase space of the intervals $\theta_j^* \in [-\pi,\pi)$  
and $\omega_j^* \in [-100,100]$, respectively, uniformly at random~\cite{Menck:2014fn}. 
When we apply such perturbation to the system, it may recover the synchronized steady state again or fall into a limit cycle (out of synchrony) after enough time~\cite{Menck:2014fn}.
For a given number of trials of random sampling of perturbation, we measure a node's $B$ value by the number of successful trials of synchronization recovery from the random samples of phase space $(\theta_j^*, \omega_j^*) \in [-\pi,\pi) \times [-100,100]$ as described above, divided by the total number of trials. We repeat this procedure for various coupling strength $K$ to systematically investigate $B$ as a function of $K$.

In terms of synchronization stability, larger values of $B$ indicate that the system endures a wider range of phase and angular frequency disturbance, which means that the node is more stable against certain perturbations. 
On the other hand, a node with $B \simeq 0$ is almost always unable to recover the synchronized steady state, even from small perturbation. 
Each node has its own transition pattern of the basin stability $B$, but in general, $B$ is usually an increasing function of the value of coupling strength $K$~\cite{Menck:2014fn, BasinStability_NJP}. 
One can intuitively understand the larger stability at larger $K$ values because larger coupling strength enhances the interaction between nodes and such a tightly coupled system tends to absorb perturbation via their  mutual collective robustness.
However, it is not always true. 
The counterintuitive phenomena of suddenly decreased $B$ have been witnessed from seemingly quite stable cases with $B \simeq 1$ even when $K$ is increased~\cite{Kim:2016kd}. 
This phenomenon, therefore, leaves characteristic peaks of $B$ as a function of $K$, and implies that the power grid may suddenly become unstable unexpectedly, even with larger values of coupling strength. 

In order to investigate the stability of power-grid nodes systematically in terms of basic building blocks or motifs constituting power grids, we conduct the basin stability analysis on nodes in small graph structures from simple chains up to the enumerative set of the four graphlets~\cite{Sarajlic:2016wx}. 
One may consider a national level of power grid as a single large network.
In reality, however, small-size micro grids, loosely isolated from the national grid, are also important, e.g., 
the local power plants in industrial complexes or highly populated areas supply the regional power demand.  
A more specific example is the power plants for renewable energy sources such as solar photo-voltaic, which supply electricity to keep the local power grid alive, even when the main power grid becomes shut down for some reasons. 
These small parts of the power grid operate independently from the main power grid in the form of micro power grids, which is called ``islanding''~\cite{Mureddu:2016dw}. 
Understanding the synchronization behavior and stability of these islands is important even in the perspective of the entire power grid stability, because the islands should maintain the local functioning of the power grid even when the blackout in the global scale occurs~\cite{Yang:2017io,Mureddu:2016dw}. Therefore, studying a set of small graph configurations is a good start to explore large structures~\cite{Milo:2002cg} and, at the same time, it is also important to manage the local power stability.

We compare the transition patterns of $B$ for nodes located at different topological positions in various structures and scales of power grids.
The amount of power input from the producers is assigned as $1\leq P_p \leq2$ depending on their topological properties, which will be presented later.
The power output to the consumers, the relative portion of which depends on the consumers' topological properties as well, is set for the total amount of power input and output to be balanced as $\sum^n_{i=1}P_i=0$, where $n$ is the total number of nodes.

\section{Results}
\label{sec:results}
\subsection{The Ordinary Pattern: Monotonically Increasing Stability}
\label{sec:pattern}
Let us consider a very large fully connected network as an ideal power grid, where all of the nodes are well connected so that synchronization between them never fails.  
In this ideal power grid, any perturbation to a power-grid node is instantly absorbed and the angular frequency of it is fixed at the rated value. 
Conventionally, a node of this type of ideal power grid is often replicated by an oscillator coupled with a busbar---so-called infinite busbar---and the system that can absorb an infinite amount of perturbation, analogous to the heat reservoir in the context of heat transfer in statistical mechanics~\cite{Menck:2012vc,Ji:2014ina}. 
In practice, the phase and frequency deviations of the infinite busbar are set to be constant, such that $\theta_f=0$, $\omega_f=0$ during the simulation.

\begin{figure}[h!]
\hfill\includegraphics[width=0.9\textwidth]{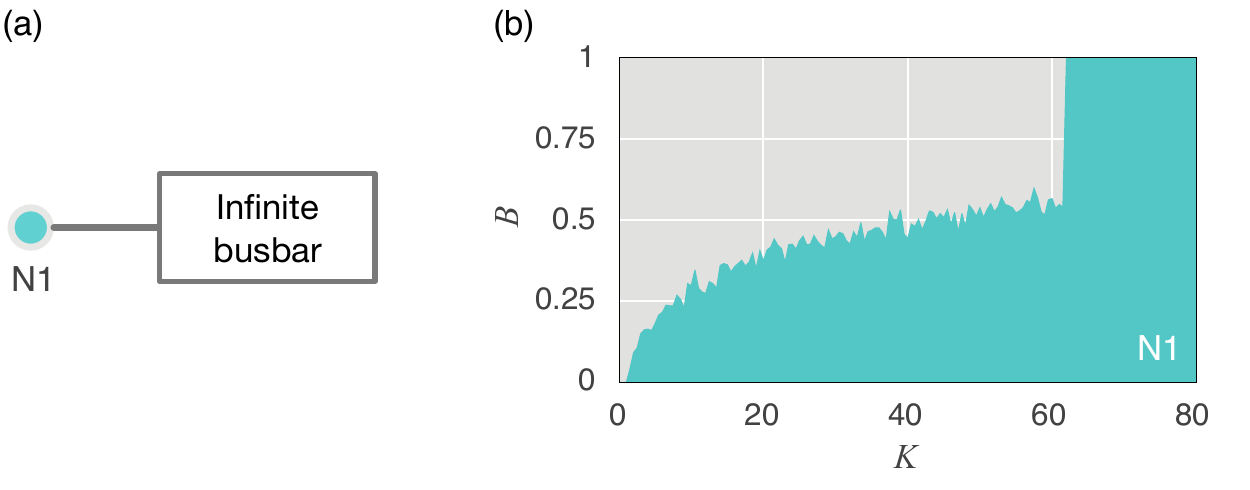}
\caption{The basin stability $B$ of a node as a function of coupling strength $K$ in the infinite busbar model. (a) A producer node is connected to a system that can absorb any perturbation from the node. (b) The node shows monotonically increasing basin stability $B$ as a function of $K$.
}
\label{fig:one_node}
\end{figure}

The transition pattern of the infinite busbar is the gradual and monotonic increase when the transmission strength increases, except for minor stochastic fluctuations from random perturbations in the phase space (see figure~\ref{fig:one_node}).
It is also reported that at $K$ values large enough to rule over the perturbation, the basin stability reaches the maximum value of unity, which means that the system can absorb any perturbation within the given range of phase and frequency space~\cite{Menck:2014fn,Kim:2016kd}. 

\begin{figure}[t!]
\hfill\includegraphics[width=0.9\textwidth]{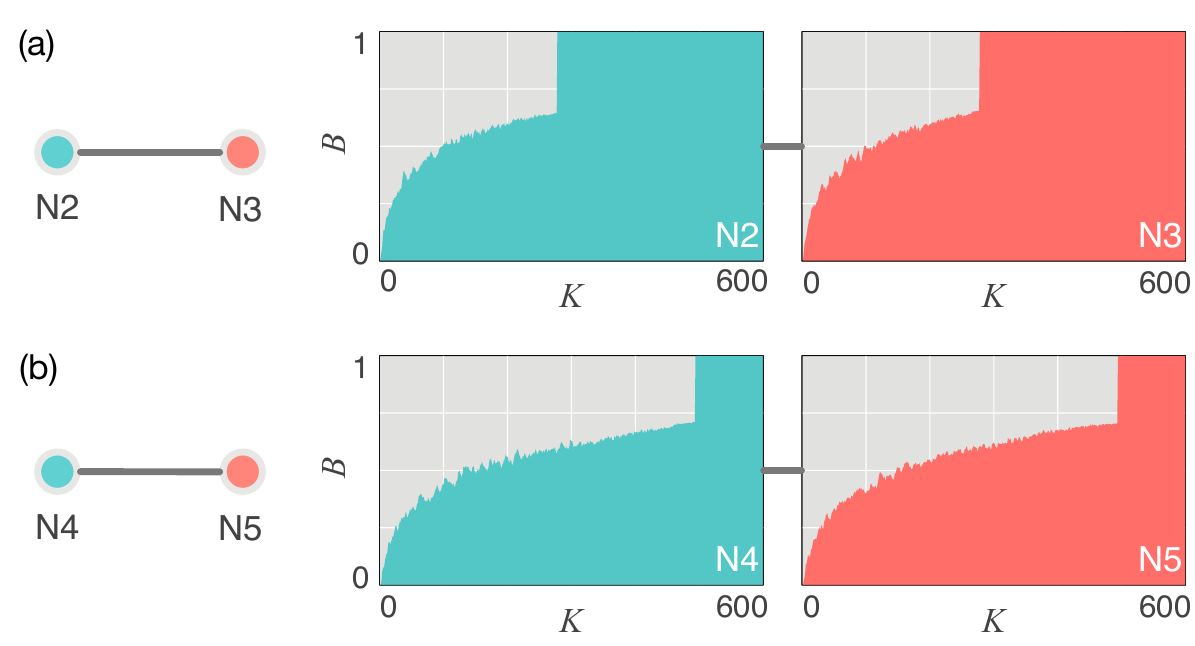}
\caption{The monotonic increase of synchronization stability in the two-node power grids, which consists of a producer node (mint on the left) and a consumer node (red on the right). The subfigures on the right show the transition of synchronization stability of each node as a function of $K$. The amount of power production $P_p=1.5$ and consumption $P_c=-1.5$ in the panel (a); $P_p=2$ and $P_c=-2$ in the panel (b).  With the increased values of $|P_i|$, the threshold $K_c$ above which $B = 1$ is also increased but the overall transition shape remains. 
}
\label{fig:two_node}
\end{figure}

The pattern of monotonically increasing basin stability as $K$ is increased and the appearance of the perfect stability $B = 1$ above a certain threshold $K_c$ also occurs in the power grid composed of two nodes as shown in figure~\ref{fig:two_node}. 
The two nodes are a power producer and a consumer with $P_p>0$ and $P_c = -P_p < 0$, respectively, for the balanced power input and output, i.e., $P_p + P_c = 0$.
The transition patterns of $B$ of the both producer and consumer are identical in this case.

The transition shape of $B$ according to $K$ sustains, although we tune the parameter values.
Changing $P$ value in the two-node case just shifts the $K_c$ after where the synchronization stability holds unity. 
Increasing the amount of input, for example, $|P|=1.5 \to 2$ from figure~\ref{fig:two_node}(a) to figure~\ref{fig:two_node}(b), the threshold for $K$ is also increased ($K_c \simeq 278 \to K_c \simeq 494$), but the overall shape of the transition pattern is not altered.
Based on this observation, the monotonic increase seems to be a universal characteristics 
of the basin stability transition, which is not always true as we present in the following.

\subsection{Emergence of Nonmonotonic Transition Pattern}
\label{sec:sync_pattern}

In contrast to the monotonically increasing transition pattern of the infinite busbar, Kim \emph{et al.}~\cite{Kim:2016kd} revealed the emergence of nonmonotonic transition patterns of the basin stability as $K$ is increased in small networks. 
The stability measured by $B$ becomes quite large for a short window of $K$, followed by a drastic decrease of $B$ with the \emph{increased} value of $K$, counterintuitively. In other words, there are local peaks of $B$ as a function of $K$, for small power-grid networks.
The nonmonotonic pattern is observable in 
small-scale power grids and is also related with the community characteristics~\cite{BasinStability_NJP}.
However, the specific underlying mechanism of the emergence of the nonmonotonic behavior is yet to be elucidated.

\begin{figure}[t!]
\hfill\includegraphics[width=\textwidth]{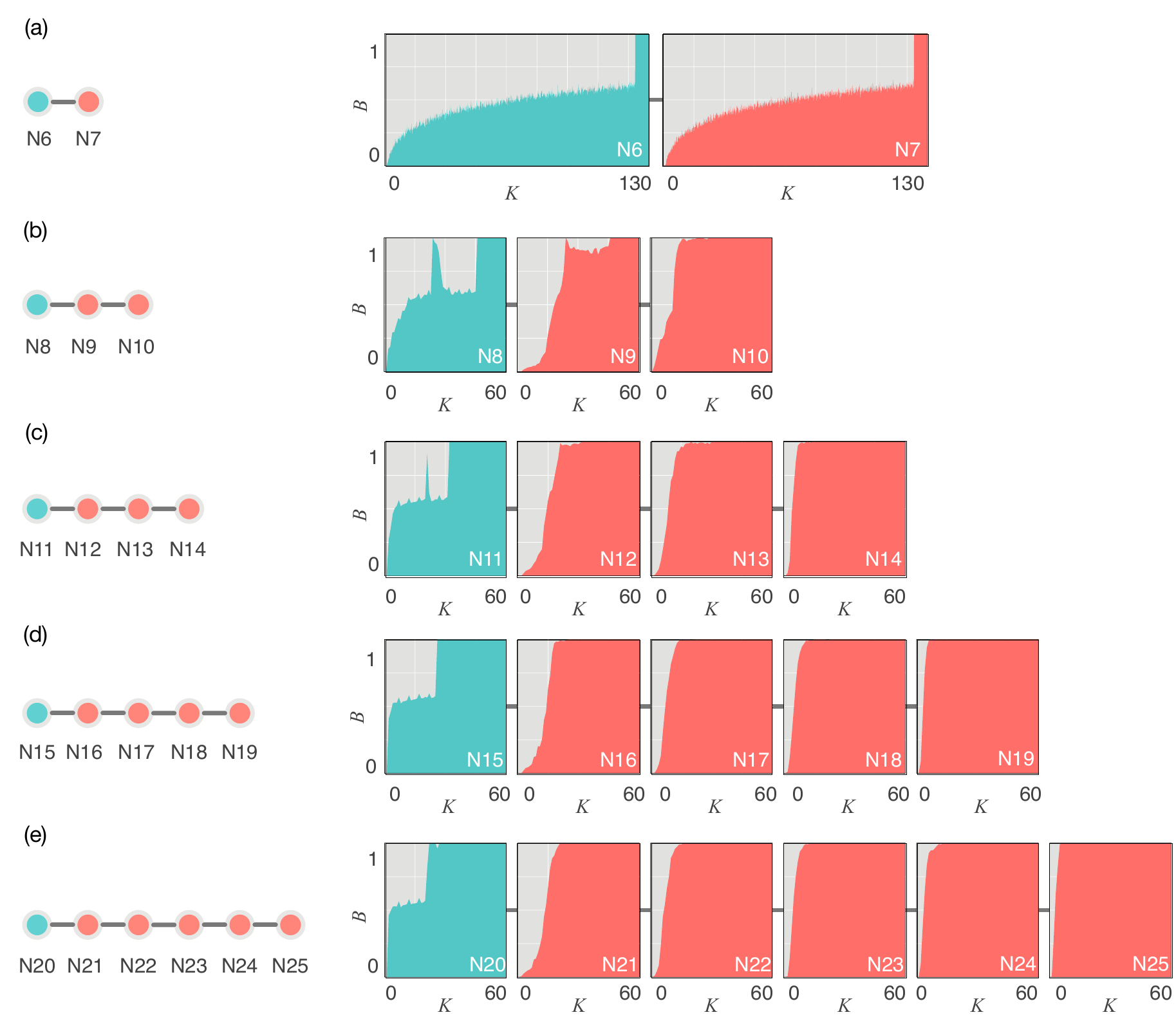}
\caption{The transition patterns of $B$ for the linear chain networks with a single producer node and different numbers of consumer nodes. The 
power input of power producers $P_p=1$ for all of the cases. The power input equally distributed to each consumer is set for the power conservation: (a) $P_c=-1/2$; (b) $P_c=-1/3$;(c) $P_c=-1/4$; (d) $P_c=-1/5$; (e) $P_c=-1/6$. Nonmonotonic transition patterns with the peak structure appear in panels (b) and (c).}
\label{fig:size-change}
\end{figure}

Our systematic numerical simulations show that the 
nonmonotonic behavior in $B$ even appears in small power-grid networks with simple topology. 
For instance, in the case of $P_p=1$, with additional power consumers to a single power plant, a power producer shows a sudden increase and 
decrease of $B$ when it is connected to a 
number of consumers (see figure~\ref{fig:size-change}). 
The peak of $B$ appears only when there are two or three consumers, as shown in figures~\ref{fig:size-change}(b) and \ref{fig:size-change}(c). 
In these cases, even if $B$ reaches the value close to unity for a certain $K$ value, the value of $B$ can suddenly droop to an intermediate value when we increase $K$ (the cases of nodes denoted by N8, N9, and N11 in figure~\ref{fig:size-change}). This peak of $B$ does not appear in the aforementioned infinite busbar and the two-node power grid. 

\begin{figure}[t!]
\hfill\includegraphics[width=\textwidth]{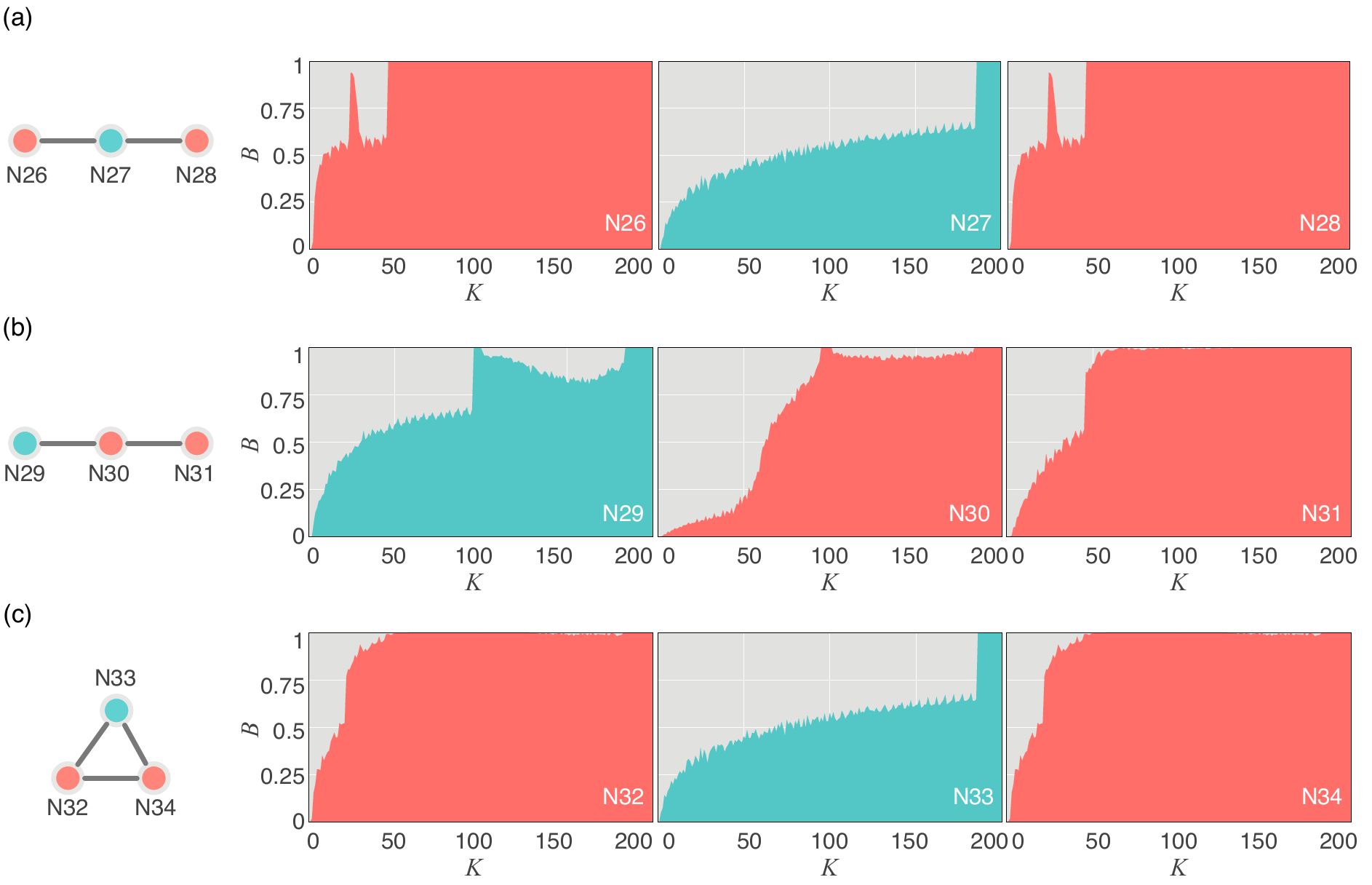}
\caption{The transition pattern of $B$ for networks composed of three nodes. Isomorphically distinct three types of topology (the networks on the left) yield various transition patterns on the right; each plot corresponds to each node in the network, respectively. The parameter values are $P_p=2$ and $P_c=-1$ 
for all of the three cases.}
\label{fig:three_node}
\end{figure}

The transition pattern with the peak is related not only to the size of the network but also to the location of power-grid components and the structural properties of the network.
For a given number of nodes, different network structures induce various transition patterns. 
For instance, connected networks composed of three nodes can belong to three different types of topology (up to isomorphism) as shown in figure~\ref{fig:three_node}, with a power producer (mint) with $P_p=2$ and two consumers (red) with $P_c=-1$. 
When the power producer is located in the center, the node denoted by N27 in figure~\ref{fig:three_node}(a), it shows a similar transition pattern to the infinite busbar or two-nodes power grids, except for the value of the critical coupling strength ($K_c \simeq 186$).
However, the consumers on the left and right sides (N26 and N28) have a sharp peak before the nodes reach $B = 1$, with much smaller $K_c \simeq 46$ than that of the central node.

In this case, a local change in a topology can affect the synchronization dynamics of the entire network.
Switching the positions of the central producer node and the left consumer node in figure~\ref{fig:three_node}(a) (N26 $\leftrightarrow$ N27) will result in the network topology in figure~\ref{fig:three_node}(b), which shows a completely different transition pattern of $B$. 
Interestingly, the level of instability in a certain range of $K \in [100,180]$ seems to become less severe from the left-end (N29, the producer) to the right-end (N31, one of the consumers). 
Moreover, when an additional edge connects both end nodes to complete the triangle (N32--N34), the peak disappears and the transition pattern become monotonic. These various transition patterns exemplify that predicting the synchronization stability is nontrivial.

\begin{figure}[b!]
\hfill\includegraphics[width=0.99\textwidth]{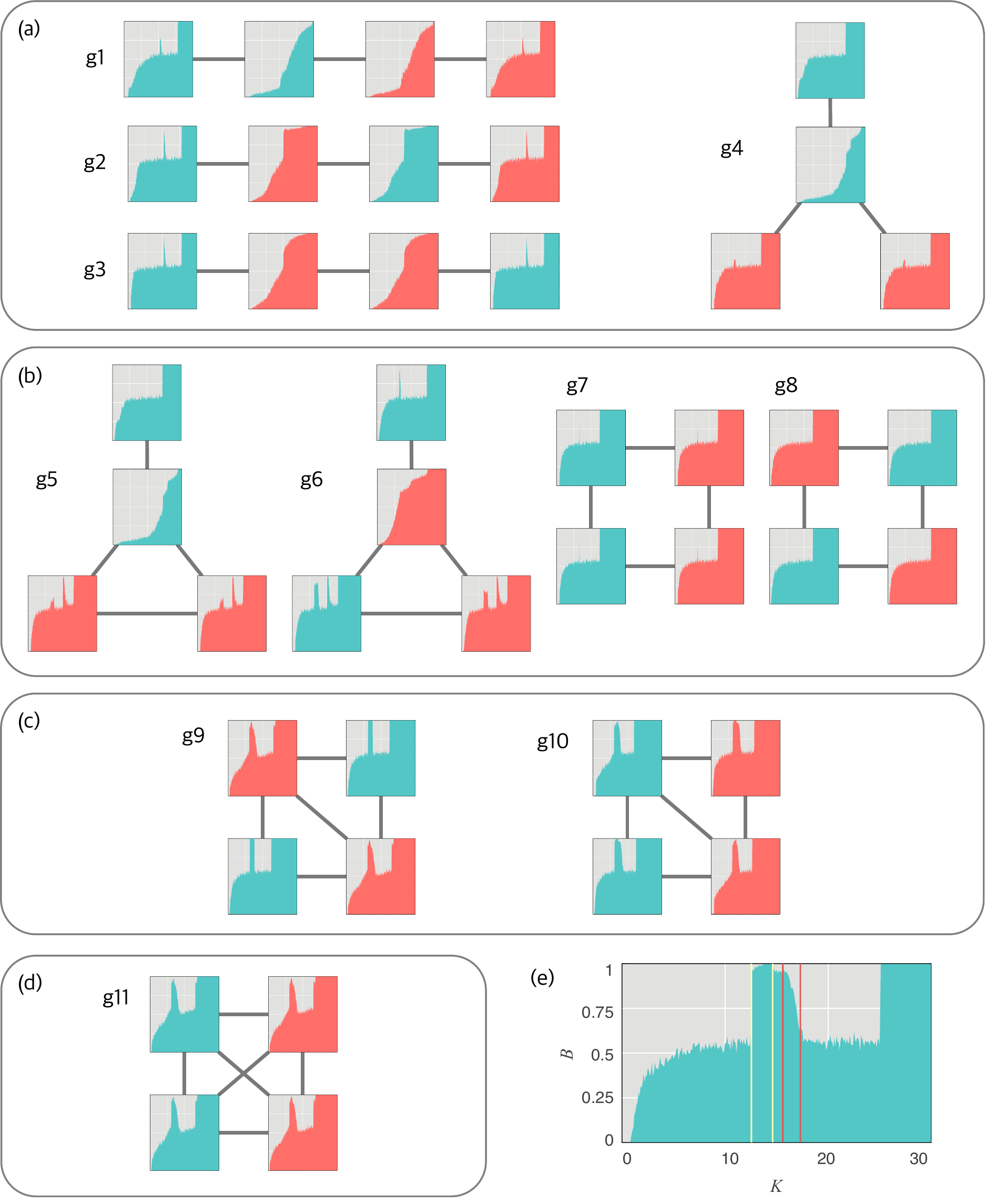}
\caption{Various transition patterns of $B$ as functions of $K$. Each network represents a topology of four-node power grids with (a) three, (b) four, (c) five, and (d) six edges.
The transition patterns of $B$ for power producers and consumers are shown in mint and red, respectively, at each of the corresponding location. All of the plots share the common legend shown in the panel (e), which is the magnified version of the bottom left node of the network denoted by g10. The vertical yellow and red lines indicate bifurcation points, which will be explained later in Sec.~\ref{sec:bifurcation}.}
\label{fig:graphlet}
\end{figure}

Such a nonmonotonic transition pattern with the peak structure becomes more common in four-node networks.
Figure~\ref{fig:graphlet} shows the various transition patterns of $B$ for four-node networks~\cite{Sarajlic:2016wx}.
In the case of connected four-node networks, 11 types of network topology are possible, when we consider the balanced network consisting of the two producers and the two consumers.

The various characteristic transition patterns of $B$ shown in the three-node power grids shown in figure~\ref{fig:three_node} are also observed in the four-node cases. 
Note that the networks where the producer and consumer sides are polarized (g1, g4, and g5 in figure~\ref{fig:graphlet}) show significant instability.
 For example, the nodes located between the producer and consumer side have a low level of $B$ for the entire range of $K$. In figure~\ref{fig:graphlet}, the two central nodes in g1 and the nodes with three neighbors in g4 and g5 are such cases. The observation is in good agreement with the report that centralized power grids are more vulnerable than distributed ones~\cite{Rohden:2014fy}.

We also find the nonmonotonic behavior in some nodes in topologies that have four nodes and five edges in figure~\ref{fig:graphlet}, including the fully-connected one (g11). 
 As a specific example, figure~\ref{fig:graphlet}(e) shows an enlarged view of the transition pattern of the bottom left node of g10 in figure~\ref{fig:graphlet}(c). 
The local peak persists with a very high value of the basin stability for $13 \lesssim K \lesssim 15$, before it decreases again for $15 \lesssim K \lesssim 18$. The basin stability reaches a final transition point to unity at $K \gtrsim 25$. 
The onset of the local peak is much sharper than its decay, which is
also observed for other topologies such as the three-node power gird in figure~\ref{fig:three_node}(b) (see the plot for N29). 

\subsection{Fine Structure of Basins of Attraction}

In the four-node networks denoted by g9, g10, and g11 in figure~\ref{fig:graphlet}(c), the power producers and consumers show very similar transition patterns in $B$ as $K$ increases. The local peak commonly appears in the range of $13 \lesssim K \lesssim 19$ for both power producers and consumers.
The overall transition shapes show a similar pattern for all nodes. 

In order to investigate this stability drooping in more detail, we investigate the variations in the instantaneous frequency of each node after the transient period, when we apply the same perturbation to a specific node. 
Since the basin of attraction in phase space changes as the coupling strength $K$ increases, the same point in phase space may not lead to the same attractor for different values of $K$. Here, we carefully choose an external perturbation to a target node such that all of the four nodes fail to recover synchronization for various values of $K$, except between $K=13$ and $14$.
In this range, all nodes show $B\simeq 1$ implying that they almost always return to the synchronized state.

\begin{figure}[t!]
\hfill\includegraphics[width=\textwidth]{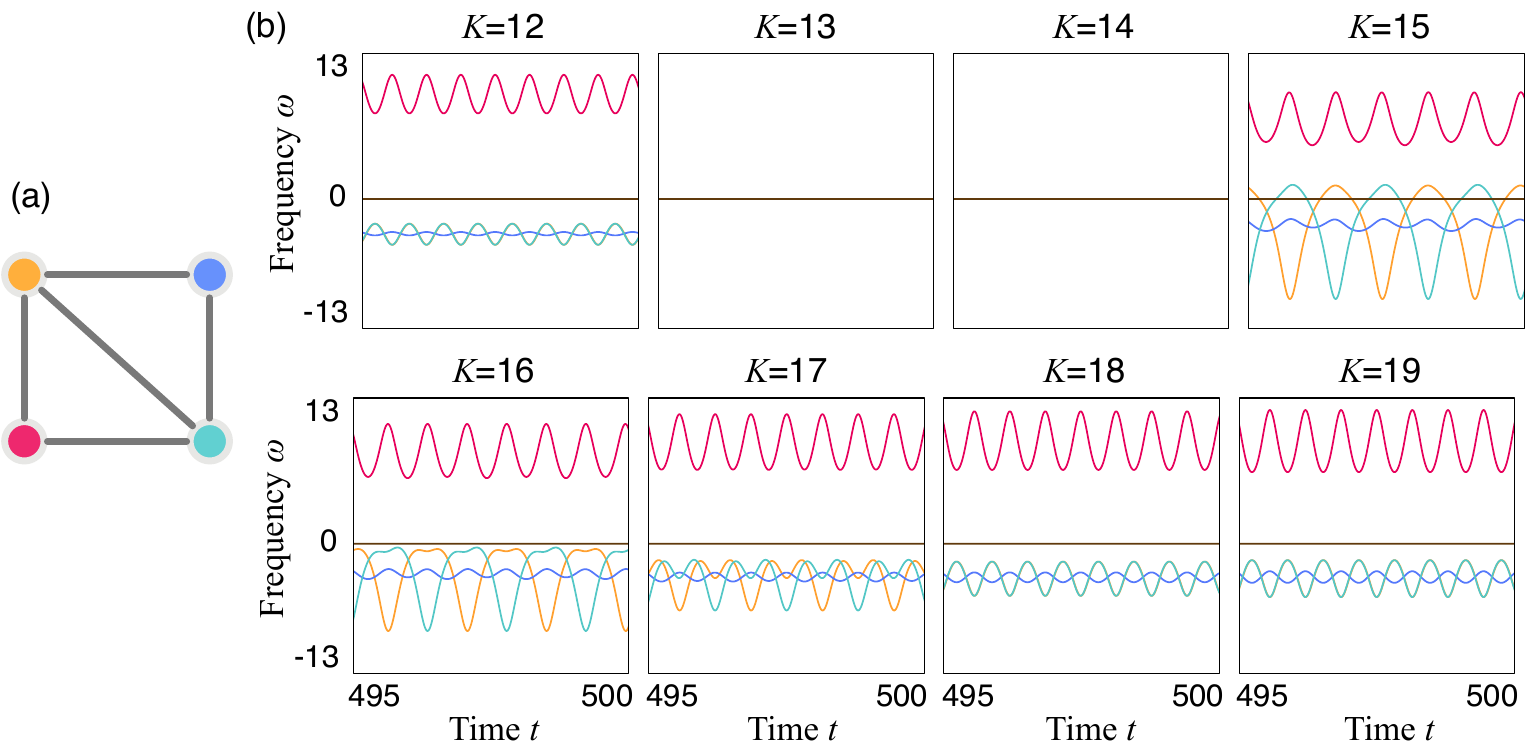}
\caption{Snapshots of the dynamics of the instantaneous frequency of each node in the four-node network denoted by g10 in figure~\ref{fig:graphlet} for various values of $K$. (a) The specific network topology: The red (bottom left) and orange (top left) nodes are the producers. The blue (top right) and mint (bottom right) nodes are the consumers. In all cases, an external perturbation is applied to the red producer node such that $\omega_{\mathrm{red}} = 16.2$, while the frequencies of all other nodes remain 0. (b) The evolution of the instantaneous frequency of each node for $K \in [12,19]$. They exhibit limit cycle behavior and do not converge to the synchronized state except for the cases of $K = 13$ and $K = 14$. The producer (orange) and consumer (mint) exhibit a 1:2 response with respect to the other nodes for $K=15$. As the coupling strength $K$ increases, a 1:1 anti-phase response is recovered. At $K\geq18$, the limit cycle is qualitatively identical to that observed at $K=12$, where the orange line and the mint line are identical. The brown guideline indicating $\omega = 0$ also represents $\sum_i \omega_i = 0$, i.e., the vanishing mean of the frequencies. 
}
\label{fig:track}
\end{figure}

As a specific example, we focus on the four-node network denoted by g10 in figure~\ref{fig:graphlet} and shown in figure~\ref{fig:track}(a). The producer nodes are located on the left (red and orange) and the consumer nodes on the right (blue and mint). The external perturbation is applied to the red producer node with $\omega_{\mathrm{red}} = 16.2$. After a sufficiently long time ($t > 495$) to minimize any transient effects, we observe the dynamics of the instantaneous frequency as shown in figure~\ref{fig:track}(b) for various $K$ values. While for $K=13$ and $K=14$ we recover a synchronized state of $\omega_i = \dot{\theta_i} = 0$ as expected, the frequency and phase of the nodes orbit a limit cycle and do not converge to the synchronized state in all other cases. More interestingly, comparing $K=12$ and $K=15$, we observe a period-doubling phenomenon in the emergent oscillations of two of the nodes---the period of the producer node (orange) and the consumer node (mint) becomes twice the period of the other nodes. When $K=15$, just after the stability drooping, these two nodes are also in anti-phase with each other, while they were in-phase and identical for $K=12$. 
However, as the coupling strength $K$ increases, we observe a reversed period-doubling and for $K = 18$ the dynamics becomes qualitatively identical to the case of a simple limit cycle at $K=12$: The orange and mint nodes have identical dynamics and they are out of phase with the other nodes, while satisfying the condition $\sum_i \omega_i = 0$. 

\begin{figure}[t!]
\hfill\includegraphics[width=\textwidth]{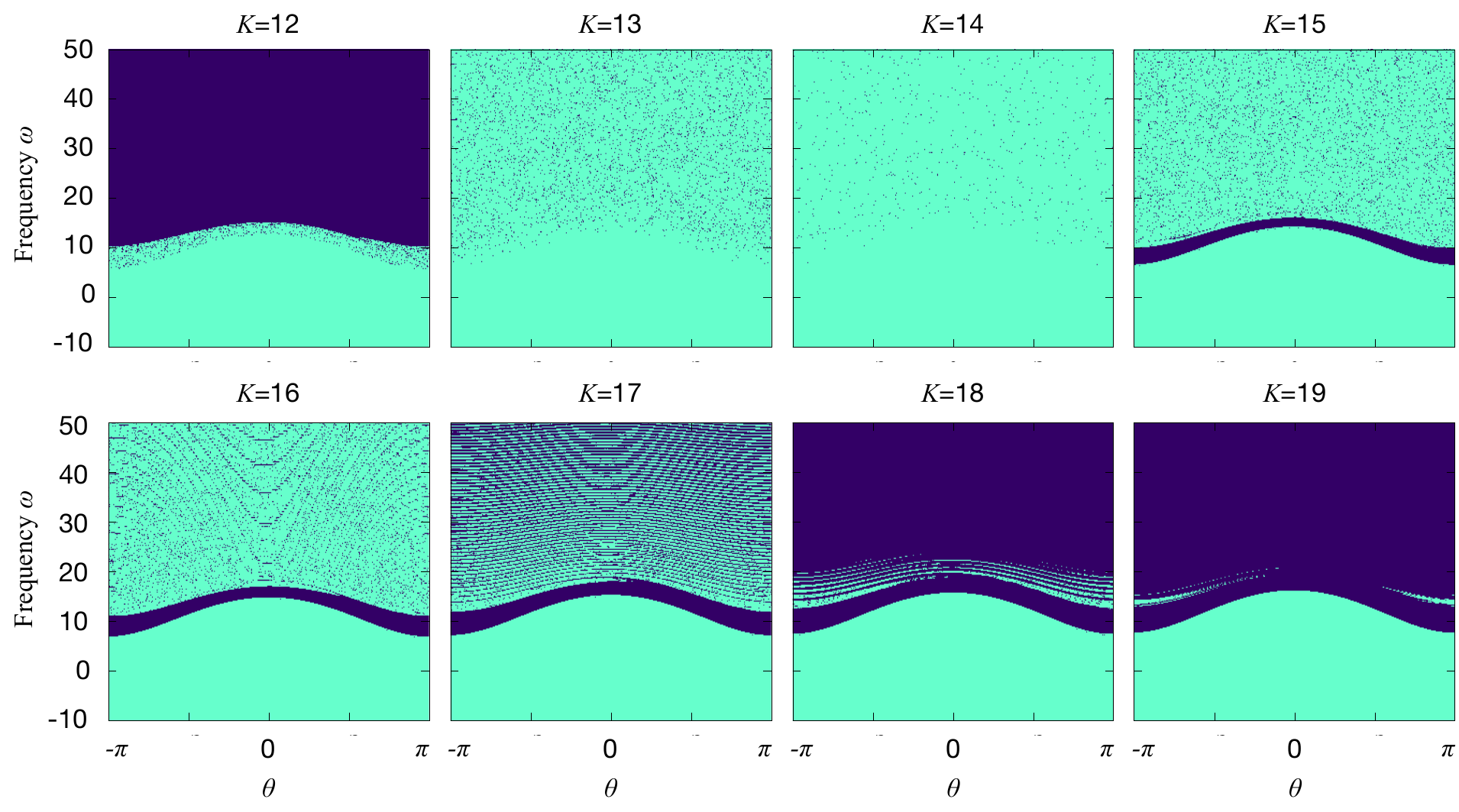}
\caption{Basins of attraction for different values of $K$ in a two-dimensional subspace of the red node in figure~\ref{fig:track}(a) (see main text for details). The light green area indicates the phase and frequency perturbation from which the system recovers synchrony, while the dark blue area indicates the phase and frequency perturbation from which the system fails to recover synchrony. Thus, the system exhibits multistability for all $K$ shown. Fractal-like fine structure patterns are observable for $13 \lesssim K \lesssim 18$, where the peak of $B$ is located. 
}
\label{fig:diagram}
\end{figure}

In order to understand the observed response of the oscillator network to the chosen perturbation better, we now explicitly explore the emergent dynamics in a more extended part of phase space $(\theta, \omega)$ for $K = 12, 13, \cdots, 19$, focusing on initial conditions $\theta \in [-\pi, \pi)$ and $\omega \in [-10, 50]$ for the red node in figure~\ref{fig:track}(a), while the initial conditions of all other nodes are $(0,0)$. Therefore, this corresponds to a 2-dimensional subspace of the full 8-dimensional phase space, but it allows for a better visualization. Figure~\ref{fig:diagram} displays eight plots of the basins of attraction for different $K$ values, where each light green point in phase space corresponds to an initial condition leading to the synchronized state.
The fraction of the light green area in each plot in figure~\ref{fig:diagram} approximately corresponds to the basin stability $B$ for the given $K$.

When the coupling strength $K=12$ in figure~\ref{fig:diagram}, more than half of the area of the shown phase space makes the system fail to recover synchronization, corresponding to the dark blue area for $\omega \gtrsim 10$, which results in $B \lesssim 0.5$ in figure~\ref{fig:graphlet}(e). However comparing figure~\ref{fig:diagram} and figure~\ref{fig:graphlet}(e) further, when $K=13$, the basin stability significantly increases and the dark blue area breaks down into small islands. This tendency continues when we increase $K$ to $K = 14$, which results in $B \simeq 1$. 
Then, the basin stability $B$ suddenly starts to decrease when we increase $K$ to $K=15$.
As shown in the plot for $K=15$ in figure~\ref{fig:diagram}, the number of the dark blue speckles increases again and in addition and more importantly a solid blue area or channel around $\omega \simeq 10$ emerges in phase space.
Interestingly, the speckled area of dark blue, corresponding to the values of perturbation from which the node fails to recover synchronization, forms a moir{\'e}-like pattern for $K=16$ and $K=17$. Eventually, around the top half of the phase space turns into the basin of attraction for the unsynchronized state (see $K=18$ and $K=19$ in figure~\ref{fig:diagram}.) These speckled patterns and the moir{\'e}-like fine structures of the basin boundary provide a more detailed insight into the stability droop.

\subsection{Multistability and Saddle-node Bifurcations of Limit Cycles}
\label{sec:bifurcation}

\begin{figure}[!t]
\hfill\includegraphics[width=\textwidth]{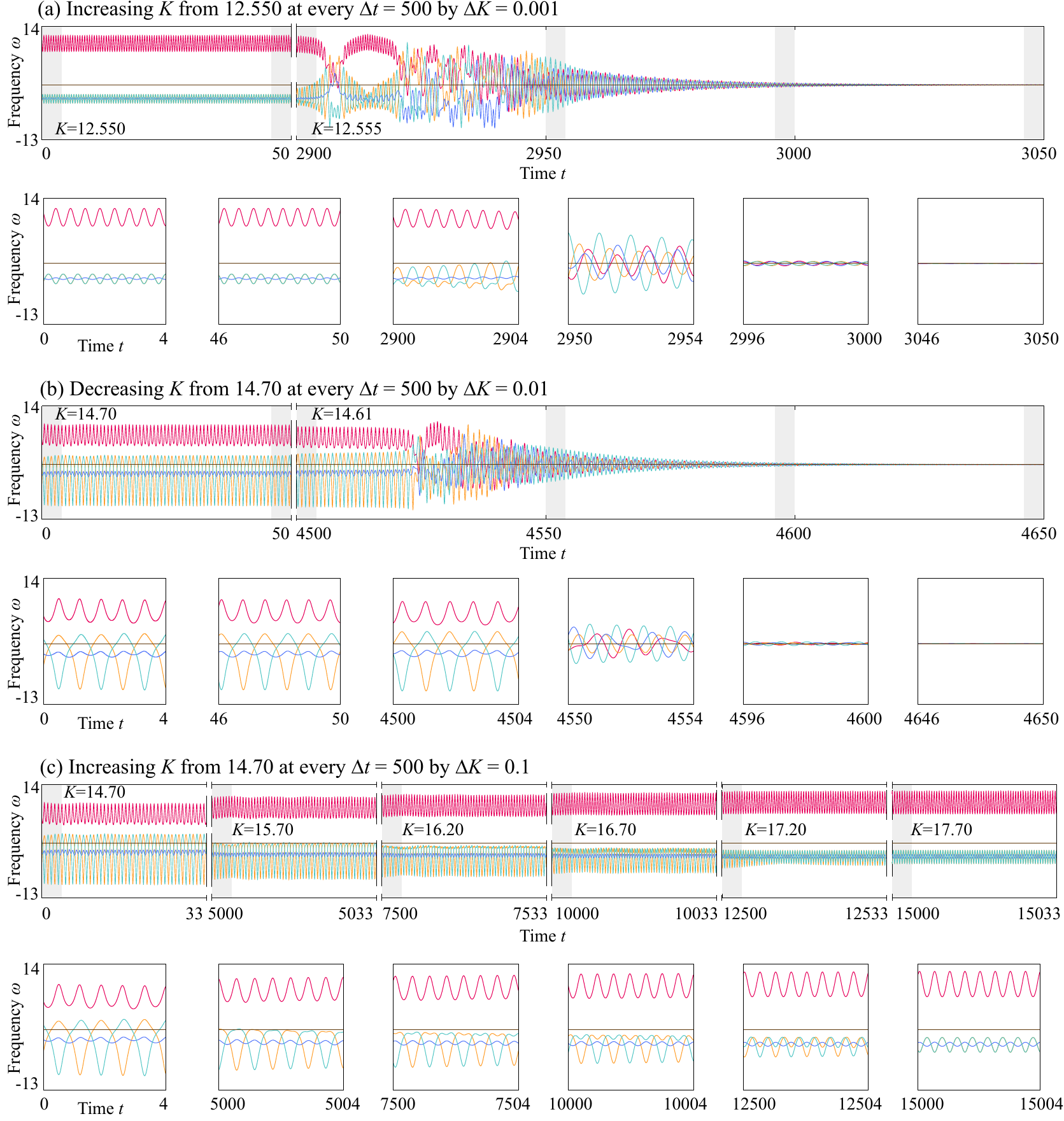}
\caption{Stability of the different types of limit cycles shown in figure~\ref{fig:track}(b) around the local peak in basin stability. Each colored line represents the instantaneous frequency of the corresponding node in figure~\ref{fig:track}(a). (a) Continuously increasing $K$ at every $\Delta t = 500$ by $\Delta K = 0.001$ starting from $K=12.55$, the simple limit cycle eventually becomes unstable giving rise to the synchronized state. The subfigures represent a magnification of the gray-shaded regions. 
(b) The same as in (a) but for the period-doubled limit cycle and continuously decreasing $K$ starting from $K=14.7$ by $\Delta K = 0.01$. This limit cycle becomes unstable at $K \simeq 14.61$ giving rise to the synchronized state.
(c) The same as in (b) but for increasing $K$ by $\Delta K = 0.1$. The period-doubled limit cycle undergoes an inverse period-doubling transition at $K \simeq 15.6$ before eventually taking on the shape of the simple limit cycle at $K \simeq 17.3$.
}
\label{fig:bifurcation}
\end{figure}

In order to analyze the observed multistability and its relationship to the changes in basin stability in more detail, we now focus on how the stability of the different limit cycles shown in figure~\ref{fig:track}(b) changes with $K$. The simple limit cycle observed for $K=12$ remains stable over a range of $K$ values. As shown in figure~\ref{fig:bifurcation}(a), at $K \simeq 12.555$ the limit cycle suddenly becomes unstable and decays to the fixed point corresponding to the synchronized state. This change in stability coincides with the onset of the local peak in basin stability (see figure~\ref{fig:graphlet}(e), where the first vertical yellow line corresponds to the instability of the simple limit cycle) indicating that the basin of attraction of the formerly stable limit cycle now almost exclusively belongs to the basin of attraction of the synchronized state.

In contrast, the period-doubled limit cycle observed for $K=15$ in figure~\ref{fig:track}(b) loses its stability for decreasing $K$. As shown in figure~\ref{fig:bifurcation}(b), this limit cycle becomes unstable for $K \simeq 14.61$ and the synchronized state is the new attractor. This bifurcation point of the limit cycles is indicated by the second yellow vertical line in figure~\ref{fig:graphlet}(e).
The instability coincides with the initial decay of the local peak suggesting a direct connection between the basin of attraction of the period-doubled limit cycle and the basin stability. Both instabilities of the two different limit cycles leading to the synchronized state as the new attractor share the property that the frequency amplitudes of the limit cycle remain finite up to the respective instability as shown in figure~\ref{fig:bifurcation}(a) and (b). This suggests that the type of bifurcation is the \emph{saddle-node bifurcation of cycles} in both cases~\cite{Strogatz}. 

Interestingly, as we increase $K$ gradually as shown in figure~\ref{fig:bifurcation}(c) the evolution of the period-doubled limit cycle follows the same pattern as observed in figure~\ref{fig:track}(b) for $K \geq 15$. Specifically, the period-doubled limit cycle first undergoes an inverse period-doubling transition around $K \simeq 15.6$ and later at $K \simeq 17.3$ a transition to the simple limit cycle, which is identical to that of $K=12$. These two transition points are indicated by red vertical lines in figure~\ref{fig:graphlet}(e), which are consistent with the more slowly decaying tail of the local peak in $B$.

\begin{figure}[!t]
\hfill\includegraphics[width=\textwidth]{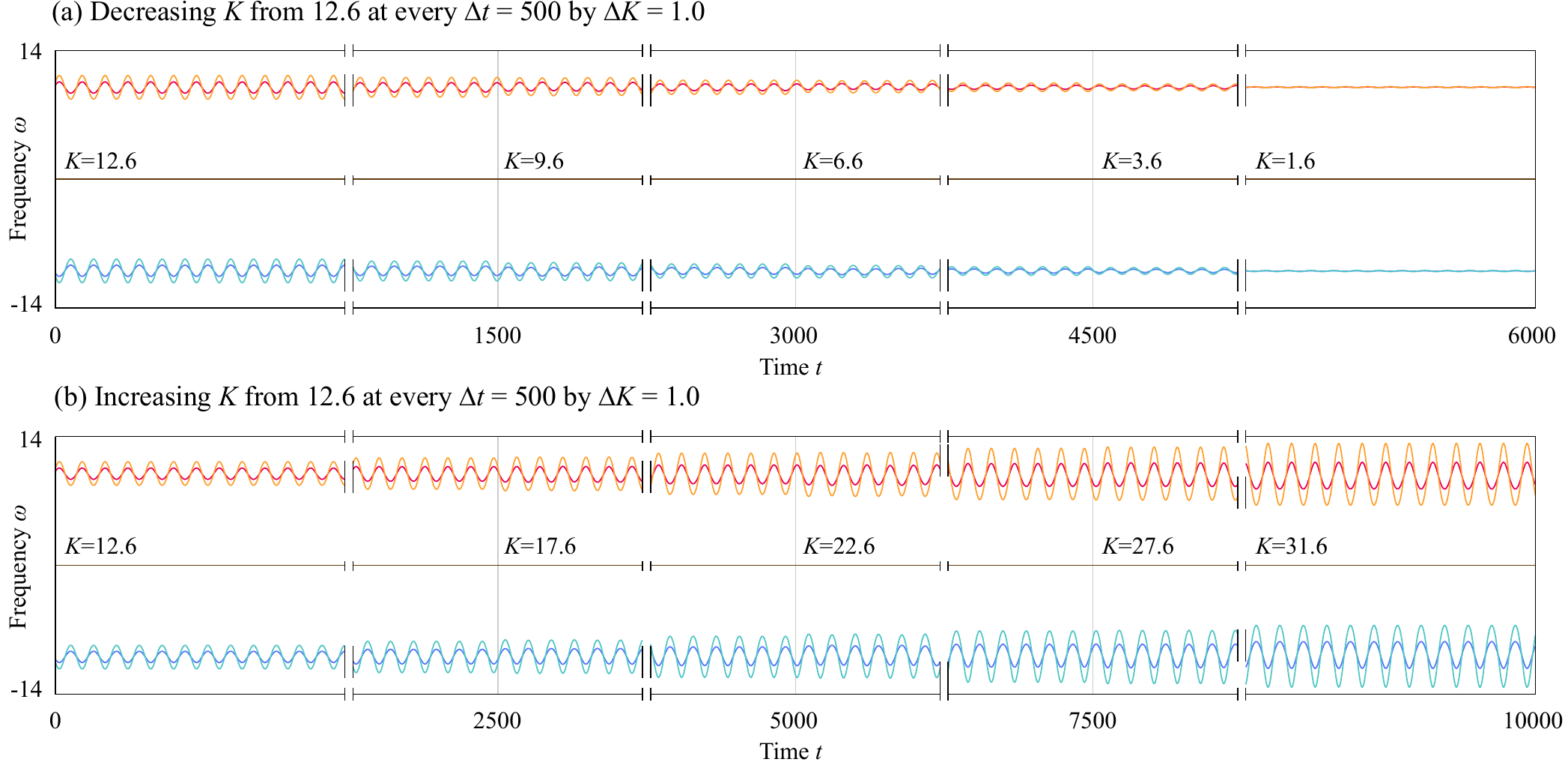}
\caption{Third observed limit cycle which remains stable over a wide range of $K$: (a) decreasing $K$ from $K=12.6$ by $\Delta K = 1.0$ and (b) increasing by $\Delta K = 1.0$. Each colored line represents the frequency of the nodes in figure~\ref{fig:track}(a).}
\label{fig:limit_cycle}
\end{figure}

Since both the the simple limit cycle and the period-doubled limit cycle are unstable between $K \simeq 12.555$ and $K \simeq 14.61$---which is consistent with the range of $K$ values in figure~\ref{fig:diagram} for which no extended blue areas are visible---the initial conditions for which synchrony is not recovered for this range of $K$ values must correspond to another attractor. Indeed, there exists another stable limit cycle as shown in figure~\ref{fig:limit_cycle}. This limit cycle has different characteristics from the others. Both producers nodes are in-phase with each other and the two consumer nodes are in-phase with each other as well. Yet, producer and consumer nodes are in anti-phase with each other. The period of the limit cycle is almost half of the former stable limit cycle in figure~\ref{fig:bifurcation}(a). In order to investigate its stability, we decrease and increase $K$ values continuously as shown in figure~\ref{fig:limit_cycle}. This limit cycle shows very stable behavior for a wide range of $K$. The period does not change with $K$, but its amplitude does.

\section{Summary and discussions}
\label{sec:summary_and_discussions}

In this study, we have investigated the transition pattern of synchronization stability $B$ as a function of coupling strength $K$ for small networks including complete graphlets up to four nodes.
Considering the characteristics of power grids that locally form a power grid islanding, this analysis enumerating up to the four graphlets can be a groundwork for power-grid networks with more complicated structures.
We find that the transition patterns of the basin stability differ for distinct types of network topology, the number of power producers and consumers, and the nodes' topological locations.

It has been conventionally and intuitively thought that the basin stability increases monotonically and reaches unity once the coupling strength is larger than a certain threshold. Our main contribution is to clearly show that the basin stability can decrease suddenly even after it becomes very high, which results in a local peak of $B$ essentially.
The emergence of such a peak has important implications because it can lead to the misinterpretation that a power-grid system is guaranteed to be stable with for all coupling strength above a certain minimal value, while in reality the basin stability may be only on the top of the peak.
In that case, the stability can suddenly decrease by even a small increment in the value of coupling strength.
Indeed, in reality, the couping strength fluctuates all the time, making it is absolutely crucial to properly understand the stability drooping phenomenon to manage the power-grid in a stable way. 

This nonmonotonic behavior is reminiscent of Braess's paradox, which in our context refers to a loss in synchrony when the capacity of a single link is a power grid is increased or a new link is added~\cite{Witthaut2012}. Indeed, as a comparison between g7 and g10 as well as between g8 and g9 in figure~\ref{fig:graphlet} shows adding a single link can lead to the emergence of a local peak in $B$.

As we showed here, the phenomenon of a local peak in synchronization stability is related to (global) bifurcations unrelated to the state of interest. Indeed, multistability and the different basins of attraction are at the core of the notion of synchronization or basin stability such that this phenomenon is not surprising from a general point of view. Yet, our study highlights the dramatic extent this phenomenon can take on in power grid graphlets with at least three nodes. The absence of the phenomenon in smaller systems suggests that the dimensionality of the system plays a significant role. More systematic and rigorous mathematical approaches will be required to investigate this in more detail, hopefully in future studies. 

\ack
This work was supported by Gyeongnam National University of Science and Technology Grant in 2018--2019 (S.H.L.), Korea--Canada Cooperative Development Program through the National Research Foundation of Korea (NRF) funded by the Ministry of Science and ICT, NRF-2018K1A3A1A74065535 (J.D. and S.-W.S.), and Basic Science Research Program through NRF-2017R1D1A1B03032864 (S.-W.S.). S.-W.S. also acknowledges the support and hospitality of the University of Calgary, Canada.

\section*{References}
\bibliographystyle{iopart-num}
\bibliography{reference}

\end{document}